\documentstyle[11pt,YSC,twoside]{article}
\def\edcomment#1{\iffalse\marginpar{\raggedright\sl#1\/}\else\relax\fi}
\reversemarginpar

\begin{document}

\title{Hot-dark matter, cold dark matter and accelerating universe}

\author{Abbas Farmany }

\affil{Young Researcher Club, Physics Department, Azad- University
of Ilam, Iran Electronic Address: a.farmany@usa.com}

\author{Amin Farmany, Mohammad Mahmoodi}

\affil{Physics Department Payamenoor University of Ilam, Iran}

\begin{abstract}
The Friedman equation is solved for a universe containing hot-dark
matter and cold dark matter. In this scenario, hot-dark matter
drives an accelerating universe no cold dark matter.
\end{abstract}

\section{Introduction}

Supernovae observations suggest that our universe is accelerating
and the dark energy contributes $\Omega_{\rm DE}\simeq 0.60-0.70$ to
the critical energy density of the present universe. In addition,
cosmic microwave background observations imply that the standard
cosmology is given by the inflation and FRW universe. On the one
hand, a non-zero cosmological constant satisfies the existence of
dark energy in the universe. On the other hand, our measurements are
claimed to apply the existence of a dark energy in the background
gravity with non-zero cosmological constant. Therefore, a typical
candidate for the dark energy is the cosmological constant. In
quantum field theory it is shown that, a short distance cutoff (UV
cutoff: $\Lambda$) is related to a long distance cutoff (IR cutoff:
$L_{\rm \Lambda}$) due to the limit set by forming a black hole.
Taking ${L_\Lambda }$ as the size of the present universe, the
resulting energy is comparable to the present dark energy. Even
though when this approach leads to the data, the description is
incomplete because it fails to explain the dark energy dominated
present universe. To resolve this situation, one is forced to
introduce another candidate for IR cutoff. One is the particle
horizon $R_{H }$that used by Fishler and Susskind. This gives
$\rho_{\rm \Lambda} \sim a^{-2(1+1/c)}$ which implies $\omega_{\rm
H}>-1/3$  however; this result corresponds to a decelerating
universe. A recent study of the x-ray emission of hot gas in a
massive cluster of galaxy has allowed astronomers to determine the
distribution of its dark matter content. The density of dark matter
appears to increase towards the center of the cluster in agreement
with cold-dark matter production. The x-ray data show that the dark
matter density increases smoothly all the way into the central
galaxy of the cluster. In the past five years, there has been
growing evidence in favor of the cold-dark matter model. The
Wilkinson Microwave Anisotropy Probe showed that normal baryon
matter only accounts for 17 percent of the matter content of the
universe; the rest being cold dark matter of unknown nature. Dark
matter particles must have the property of increasing with each
other and with normal baryon matter only through gravity. These
so-called weakly interacting massive particles are difficult to
detect and have been elusive until now. Massive neutrinos are a
possible dark matter candidate, usually referred to them as hot dark
matter because they travel at close the speed of light. Due to this
high speed, hot dark matter models of the early universe create big
structure of the size of galaxy clusters, which then fragment to
form galaxies. By contrast, the slower cold dark matter particles
cannot travel as far and so form small galaxies, which then merge to
form the bigger structures such as clusters of galaxies. Our problem
is related to answer the question: hot dark matter drives an
accelerating universe or cold dark matter. The Friedman equation is
solved for a universe contains hot-dark matter and could dark
matter. It is shown that, hot-dark matter drives an accelerating
universe no cold dark matter.

\section{Massive neutrinos as hot dark matter drive an accelerating
universe}

Now, we explain the effects of massive neutrinos (as natural
candidate for hot dark matter) to accelerating universe. The
constraint on the IR cutoff of a box with volume $V=4\pi L^3/3$ in
an effective field theory is given by,
\begin{equation}
\label{eq1}
E_{\rm \Lambda} \sim L_{\rm \Lambda}^3\Lambda^4 \le
M_{\rm S} \sim L_{\rm \Lambda} M_p^2,
\end{equation}

where $L_{\rm \Lambda} \sim \Lambda^{-2}$ is the IR cutoff scale.

Let take $L_{\rm \Lambda}$  as the size of our universe ($L_{\rm
\Lambda}=d_{\rm H}=1/H$) . The resulted energy is comparable to the
present dark energy. In this case the Friedman equation including a
matter of $\rho _m$ is given by $\rho _m =3(1-c^2)M_p^2H^2/8\pi$
which leads to an uncertainty with $w=0$. However, accelerating
universe requires that$L_{\rm \Lambda}=R_{\rm H}=a \int_0^t (dt/a)=a
\int^a_0(da/Ha^2)$. So this case is excluded. For a universe
containing cold dark matter one can take the case of the Friedman
equation $H^2=8\pi \rho_{\rm \Lambda}/3M_p^2$ leading to an integral
equation $HR_{H}=c$ and taking the form
\begin{equation}
\label{DEP}c\frac{d}{dt}\Big(\frac{H^{-1}}{a}\Big)=\frac{1}{Ha^2}.
\end{equation}
The energy density is  $\rho_{\rm \Lambda}\sim a^{-2(1+1/c)}$. This
case is still a decelerating phase because the moving Hubble
scale($H^{-1}$/a) is increasing with time. An accelerating universe
is satisfied by
\begin{equation}
\label{eq2}
 \ddot{a}>0 \leftrightarrow
\frac{d}{dt}\Big(\frac{H^{-1}}{a}\Big)<0 \leftrightarrow
\omega<-\frac{1}{3},
\end{equation}
So, in an inflationary universe the changing rate of $H^{-1}/a$ with
respect to $a$ is always negative. Let us solve the Friedman
equation in a universe containing hot-dark matter. In the early
universe, hot dark matter creates the big structure of universe.
Massive neutrinos (as natural candidate for hot dark matter) travel
at close to the speed of light. According to relativity theory, in
the viewpoint of an observer (due to the light speed), the time
expansions will occur. For this reason, we must take the integral
bound of Fishler-Susskind integral from,
\begin{equation}
\label{eq3} $$ $\lambda$ $=R_{\rm h}=a \int_0^{t} (dt/a)=a
\int_0^{a}(da/Ha^2)$ $$
\end{equation}
to,
\begin{equation}
\label{eq4} $$ $\Lambda=R_{\rm h}=a \int_t^{\infty} (dt/a)=a
\int_a^{\infty}(da/Ha^2)$ $$
\end{equation}
Using an integral form of Friedman equation of $HR_{H}$=c, we
obtain,
\begin{equation}
\label{eq5}
c\frac{d}{da}\Big(\frac{H^{-1}}{a}\Big)=-\frac{1}{Ha^2}.
\end{equation}
This leads to  $\rho_{\rm \Lambda}\sim a^{-2(1-1/c)}$ with
$\omega_{\rm h}= -1/3(1+2/c)$ that corresponds to an accelerating
universe.
\section{Conclusion}

Dark particles must have the property of increasing with each
other and with normal baryon matter only through gravity.
So-called weakly gravity interacting massive dark particle are
difficult to detect and have been elusive until now. Massive
neutrinos are a possible dark particle candidate, usually referred
to them as hot dark matter because they travel at close to the
speed of light, due to this light speed, hot dark matter models of
the early universe create big structure of the size of galaxy
clusters, which then fragment to form galaxies. By contrast, the
slower cold dark matter particles cannot travel as far and forms
the small galaxies first, which then merge to form the bigger
structures such as clusters of galaxy. We showed that, the hot
dark matter (neutrinos) drives an accelerating universe.

\acknowledgments

Farmany would like to thanks Mahmood Rostaminia, Safa Farrokhi,
for financial supports and M. S. Pournajaf, M. Farhat for
hospitality.

\end{document}